\documentclass{kapproc} 
\usepackage{procps}
\usepackage[dvips]{graphicx}
\upperandlowercase
\kluwerbib 

\begin{document}

\articletitle{The IMF in Starbursts}

\author{Bruce G. Elmegreen}
\affil{IBM Research Division, T.J. Watson Research Center\\P.O.
Box 218,Yorktown Heights, NY 10598, USA}
\email{bge@watson.ibm.com}

\begin{abstract}
The history of the IMF in starburst regions is reviewed. The IMFs
are no longer believed to be top-heavy, although some superstar
clusters, whether in starburst regions or not, could be. General
observations of the IMF are discussed to put the starburst results
in perspective. Observed IMF variations seem to suggest that the
IMF varies a little with environment in the sense that denser and
more massive clusters produce more massive stars,
and perhaps more brown dwarfs too, compared to intermediate mass
stars.
\end{abstract}


to be published in ``Starbursts: from 30 Doradus to Lyman Break
Galaxies,'' held at Institute of Astronomy, Cambridge University, 
UK, September 6-10, 2004. Kluwer Academic Publishers, edited by
Richard de Grijs and Rosa M. González Delgado.

\section{Introduction: History of Starburst IMFs}

Early starburst observations suggested the luminous mass from
young massive stars is comparable to the dynamical mass from the
rotation curve (see reviews in Telesco 1988; Scalo 1990; Zinnecker
1996; Leitherer 1999). This implied there was a deficit in low
mass stars. The nearby starburst galaxy, M82, was one of the best
cases. Rieke et al. (1980, 1993) modeled M82's gas mass, luminous
star mass, rotation curve mass, 2.2 $\mu$m flux, Lyman continuum
flux, CO index for supergiants, and the Br$\alpha$/Br$\gamma$
ratio, giving an extinction $A_V=25$ mag. They concluded that only
the usual IMF models with lower mass limits $M_L>3-6$ M$_\odot$
worked. Kronberg et al. (1985) used the Lyman continuum flux from
radio emission in M82 to determine the star formation rate, and
used the total gas mass combined with an efficiency estimate to
conclude that the IMF is top-heavy compared to the Miller \& Scalo
(1979) IMF. Bernlohr (1992) fit the same M82 properties as Rieke
et al., plus the heavy element abundance and FIR line ratios, and
concluded that either the IMF slope is shallower than the Scalo
(1986) IMF by 1, or there is a lower mass cutoff greater than
1.5-2 M$_\odot$. Independent models of M82 by Doane \& Mathews
(1993), emphasizing the SN rate and the total dynamical mass, led
to the same lower cutoff even for the Salpeter IMF, which is
shallower than both the Miller \& Scalo and Scalo functions.

Early observations obtained truncated IMFs in other starbursts
too. Wright et al. (1988) found $M_L>3-6$ M$_\odot$ from low M/L
ratios in 12 starburst galaxies assuming a Miller-Scalo IMF. For
the merger remnant NGC 3256, Doyon, Joseph \& Wright (1994)
modeled the Br$\gamma$ equivalent width, HeI $\lambda2.06$
$\mu$m/Br$\gamma$, CO index, and N(Lyc)/LIR ratio with different
IMFs and star formation histories. The HeI $\lambda2.06$
$\mu$m/Br$\gamma$ index suggested an upper mass limit $M_U=30$
M$\odot$, the rotation curve and CO molecular cloud observations
gave $M_{\rm gas}\sim M_{\rm total}$, and all constraints gave
either an IMF slope shallower than the Salpeter slope by $\sim0.5$
or $M_L>3$  M$\odot$.   For the merger remnant UGC 8387, Smith et
al. (1995) suggested $M_L>8$ M$_\odot$ for a Miller-Scalo IMF
using the low ratio of 1-500 $\mu$m flux to Ly$\alpha$, as
obtained from the 5 GHz flux. Smith, Herter \& Haynes (1998) got
the same result from IR excesses in 20 starburst galaxies,
suggesting that the IMF slope for $M>10$ M$_\odot$ was
$-2.7\pm0.2$ ($\Gamma=-1.7$, see below), shallower than the
Miller-Scalo slope of $-3.3$ ($\Gamma=-2.3$) in this mass range,
but steeper than Salpeter ($-2.35$; $\Gamma=-1.35$).

At about the same time as these observations were suggesting the
IMF was top-heavy in starburst galaxies, several other
observations suggested it was normal. Devereux (1989) observed 20
nearby starbursts like M82, and using 2.2 $\mu$m, FIR, and
estimates for the central dynamical masses, found acceptable fits
to the Miller-Scalo IMF from $0.09-30$ M$_\odot$ (and the
Miller-Scalo function is the least top-heavy of the main IMF
models). He also suggested that extinction corrections in M82 made
by others were too high, and this made it appear like M82 had a
truncated IMF when really it didn't. Satyapal (1995) indeed found
low extinction in M82 from Pa$\beta$/Br$\gamma$, which gave
$A_V=2-12$ mag compared to 25 mag in Rieke et al. (1980). Satyapal
then got a K-band luminosity 3 times lower than Rieke et al., and
saw no need for IMF truncation. Satyapal (1997) also found an age
gradient in the center of M82 and fit an IMF with the Salpeter
slope from $0.1-100$ M$\odot$, accounting for only 36\% of the
dynamical mass.

In other observations of starburst IMFs, Schaerer (1996) applied
evolutionary models to the WR/O star ratios and found a Salpeter
IMF slope, although $M_L$ could not be determined. Stasi'nska \&
Leitherer (1996) modeled the emission line spectra of giant HII
region and starburst galaxies, having a factor-of-ten range in
metallicities, and also found a Salpeter IMF up to 100 M$_\odot$,
with no information about the lower mass limit. Calzetti (1997)
modeled multiwavelength spectroscopy and broad-band infrared
photometry of 19 starburst galaxies to derive reddening values and
found a general consistency with the Salpeter IMF between 0.1 and
100 M$_\odot$.

Finally, in returning to M82, Förster Schreiber et al. (2003)
modeled it with 25 pc resolution using near-IR integral field
spectroscopy and mid-IR spectroscopy. An upper mass limit to stars
greater than 50 M$_\odot$ was derived from the $L_{\rm bol}/L_{\rm
Lyc}$ and [NeIII]/[NeII] ratios; short decay times were observed
for star formation locally (1-5 My), and the models were
insensitive to the shape or slope of IMF at intermediate to high
mass. An IMF turnover somewhere below 1 M$_\odot$ was concluded.

There were also suggestions that top-heavy IMFs would cause
problems with stellar populations or metallicities. Charlot et al.
(1993) suggested that an inner-truncated IMF would produce a very
red population of red giants, without the corresponding main
sequence stars, after the turnoff age reaches the stellar lifetime
at the truncation mass. Wang \& Silk (1993) suggested that the
truncated model gives an oxygen abundance that is too high when
the star formation process is over.  These red populations or
elevated oxygen abundances have not been observed.

At the present time, the observations suggest that the Salpeter
IMF with a lower mass flattening somewhere between 0.5 M$_\odot$
and 1 M$_\odot$ is a reasonable approximation to the IMF in large
integrated regions of starburst galaxies.

\section{Local IMFs}

\subsection{The Field}

Starburst IMFs are useful for understanding star formation only in
comparison to local IMFs or IMFs in non-starburst regions, where
many of the star formation processes can be observed in more
detail. There are many such observations of non-starburst IMFs, as
reviewed in the conference proceedings {\it The Stellar Initial
Mass Function} edited by ed. Gilmore, Parry \& Ryan (1998) or in
Chabrier (2003). We give a brief summary here.

In 1955, Salpeter showed that if 10\% of a star's mass is
converted into Helium on the main sequence, if the star formation
rate in the local Milky Way disk is constant over time, and if the
present day mass function is that given by the available catalogs
(some of which dated back several decades -- even into the
1920's), then the mass function of stars at birth has a slope of
$\Gamma=-1.35$ on a log-log plot.

More recent derivations of this field star IMF generally give
steeper slopes. Miller \& Scalo (1979) fitted the observations to
a log-normal IMF, which has about the Salpeter slope near one
solar mass, but an increasingly steep slope toward higher masses,
reaching $\Gamma\sim-2.3$. Scalo (1986) found a field star IMF
with a slope between $-1.5$ and $-1.7$ in the range from 1 to 10
M$_\odot$, and a slightly shallower slope, between 1.35 and 1.5,
at higher mass. Rana (1987) derived a field IMF with somewhat
different data than Scalo (1986) and found $\Gamma = -1.8$ for
$M>1.6$ M$_\odot$.

The difference between these field star IMFs and Salpeter's IMF is
significant: for a slope difference of 0.5, the number of high
mass stars between 10 M$_\odot$ and 100 M$_\odot$ compared to the
number of intermediate mass stars between 1 M$_\odot$ and 10
M$_\odot$ is three times larger in the Salpeter IMF than in the
others. This excess factor of 3 for nearby high mass stars can
easily be ruled out. However, Salpeter did not have observations
that extended to the high mass range. In the region of overlap,
which is near 1 M$_\odot$, the modern field star IMF slope is
comparable to Salpeter's value.  The big question for starburst
regions is how the IMF near $1$M$_\odot$ extrapolates to OB stars.

IMF measurements outside starburst regions give a wide range of
slopes. Parker et al. (1998) did photometry on 37,300 stars in the
LMC \& SMC, and found slopes concentrating near $\Gamma=-1.0,
-1.6$, and $-2.0$ for the Davies, Elliot \& Meaburn (1976) HII
regions, and $\Gamma=-1.80\pm0.09$ for all the field stars,
considering only stars with $M>2$ M$_\odot$. The IMFs near the HII
regions are probably too shallow as a result of inadequate
corrections for background and foreground stars (Parker et al.
2001), but the field star IMF appears to be free of this
systematic effect. Note that the statistical accuracy is very high
for this measurement.

Massey et al. (1995) and Massey (2002) surveyed the remote fields
in the LMC and SMC, defining these to be regions more than 30 pc
from a Lucke \& Hodge (1970) or Hodge (1986) association. The
survey was complete down to 25 M$_\odot$ and included 450 stars,
which should give a statistical uncertainty of
$\Delta\Gamma\sim\pm0.15$ (Elmegreen 1999a). By assuming a
constant star formation rate over the last 10 My, Massey et al.
found $\Gamma$ significantly steeper in the remote field than in
clusters, having a value between $-3.6$ and $-4$.

One could imagine several systematic effects that make this slope
artificially steep. First note that runaway O stars could not do
this, because the field has too few O stars compared to
intermediate mass stars. Other likely processes could do it,
however: (1) Selective evaporation of cluster envelopes into the
field, considering that some cluster envelopes have $\Gamma\sim4$
already (de Grijs, et al. 2002). (2) Greater migration into the
field of the longer-lived, low-mass stars compared to high mass
stars. (3) Greater self-destruction of low pressure clouds in the
field by OB star formation compared to high-pressure clouds in
associations (Elmegreen 1999a).  Hoopes, Walterbos \& Bothun
(2001) also explained the steep mass function required for diffuse
interstellar ionization in nearby galaxies with the differential
drift of low-mass stars into the field. Tremonti et al. (2002)
found a Salpeter IMF for clusters and a steeper IMF for the field
in the dwarf starburst galaxy NGC 5253, and explained this
difference as a result of cluster dispersal after 10 My, when the
most massive stars have disappeared.

Another observation of a systematically steep IMF was by Lee et
al. (2004). They fit the high M/L in low surface brightness
galaxies with population synthesis models requiring low
metallicity, recent (1-3Gy) star formation, and a steep IMF:
$\Gamma=-2.85$ from $0.1-60$ M$_\odot$.  These galaxies have a low
pressure like the extreme field regions in Massey et al..

The Taurus region in the nearby field may have a steep IMF at high
mass too (Luhman 2000).  The pre-stellar condensations there are
peculiar anyway, showing more extended structures, like isothermal
spheres, than those in Perseus or Ophiuchus which appear truncated
(Motte \& André 2001).

\subsection{Clusters}

The IMF in the Sco-Cen OB association is best fit with a slope of
$\Gamma\sim-1.8$ (Preibisch et al. 2002). Another Galactic region,
W51, has 4 subgroups, all with $\Gamma\sim-1.8$ at intermediate to
high mass, but two of the subgroups have a statistically
significant excess by factor of $\sim3$ of stars in the highest
mass bin ($\sim60$ M$_\odot$; Okumura et al. 2000). There are
other anomalies like this too. Scalo (1998) suggested that the IMF
varied significantly from cluster to cluster, but Elmegreen
(1999a) and Kroupa (2001) showed that most of these variations
could be statistical in origin, given the small number of stars
usually observed.

Most clusters have Salpeter IMFs. A good example is the R136
cluster in the 30 Dor region of the LMC. The slope is
$\Gamma\sim1.3-1.4$ out to stellar masses greater than $100$
M$_\odot$ (Massey \& Hunter 1998). In addition, h and $\chi$
Persei (Slesnick, Hillenbrand \& Massey 2002), NGC 604 in M33
(González Delgado \& Pérez 2000), NGC 1960 and NGC 2194 (Sanner et
al. 2000), NGC 6611 (Belikov et al. 2000) and many other clusters
have Salpeter IMFs (e.g., Sakhibov \& Smirnov 2000; Sagar, Munari,
\& de Boer 2001). Massey \& Hunter (1998) concluded that the
Salpeter IMF occurs in star-forming regions spanning a factor of
200 in density.

Whole galaxies are often observed to have the Salpeter slope too.
There were many studies in the 1990's using H$\alpha$ equivalent
widths, spectro-photometry, metallicity, and galaxy evolution
models (see review in Elmegreen 1999b). More recently, Baldry \&
Glazebrook (2003) derived the Salpeter IMF from the cosmic star
formation rate, Rejkuba, Greggio \& Zoccali (2004) got it for the
halo of NGC 5128 (Cen A), and Pipino \& Matteucci (2004) fit the
photochemical evolution of elliptical galaxies with a Salpeter
IMF.

If whole galaxies have the same average IMF as clusters, and if
most stars form in clusters, which is believed to be the case
(Lada \& Lada 2003), then the mass of any star cannot depend on
the cluster mass. That is, any type of star can form in any type
of cluster (as long as the cluster mass is larger than the stellar
mass).  If this were not the case, then the summed IMFs would
differ from the cluster IMFs. For example, if low mass clusters
were able to form only low mass stars, and high mass clusters
formed all types of stars, as observed, then the sum of the low
and high mass clusters would produce far more low mass stars than
each cluster's IMF (Elmegreen 1999b).

\section{Top heavy IMFs in Super Star Clusters}

Some super star clusters (SSC) apparently have ``top heavy'' or
``bottom light'' IMFs.  Sternberg (1998) found a high L/M ratio in
NGC 1705-1 and concluded that either $|\Gamma|<1$ or there is an
inner-mass cutoff.   Smith \& Gallagher (2001) got a high L/M in
M82F and inferred an inner cutoff at 2-3 M$_\odot$ for
$\Gamma=-1.3$; they also confirmed inner truncation for NGC 1705-1
found by Sternberg.  Alonso-Herrero et al. (2001) observed a high
L/M in the starburst galaxy NGC 1614, suggesting a top-heavy IMF.
McCrady et al. (2003) found that MGG-11 in M82 is deficit in low
mass stars.  Mengel et al. (2002) found the same for the antennae,
NGC 4038/9, and noted that the clusters in the high pressure
regions had more normal IMFs, as if the Jeans mass were lower
there.

Other SSCs have normal IMFs, however. This is the case for NGC
1569-A (Ho \& Filippenko 1996; Sternberg 1998), NGC 6946 (Larsen
et al. 2001), and M82: MGG-9 (McCrady et al. 2003).

Measuring the IMF in SSCs is subject to many uncertainties. It
requires observations of the velocity dispersion and radius to get
the mass, and observations of the luminosity. One problem is that
$\Delta v$ can vary inside a cluster (i.e., it may not be
isothermal -- e.g. NGC 6946) and it is often measured with large
uncertainties. The value of R is uncertain too if the core is
unresolved or the outer part of the cluster is blended with field
stars. The luminosity is uncertain because of possible field star
blending. Mass segregation makes the IMF vary with radius (de
Grijs et al. 2002), so the cluster colors vary with radius, giving
another uncertainty about the core radius. The average IMF depends
on where the outer cutoff is placed.  The cluster could also be
evaporating or out of radial equilibrium, in which case the usual
expressions for cluster mass in terms of $\Delta v$ and $R$ do not
apply. Several SSCs are observed to have sub-clusters inside their
halos, giving irregular and asymmetric light profiles.

\subsection{Implications}

These observations suggest a correlation between $\Gamma$ and star
formation density. In the extreme field, $\Gamma \sim-4$; in low
surface brightness galaxies, $\Gamma\sim-2.85$; in the Milky Way
and LMC fields, $\Gamma\sim-1.8$; in many clusters,
$\Gamma\sim-1.35$; in some SSC, $|\Gamma|<1.35$ or there is an
inner-mass truncation, and in starburst regions as a whole,
$\Gamma\sim-1.35$ with or without inner-truncation (this is
uncertain). We should probably remove the local field from this
list because it is a mixture of dispersed clusters and cluster
envelopes integrated over time; both the mixing process and the local
star formation history are uncertain. Aside from this, the trend
suggests significantly denser regions have slightly shallower IMF
slopes at intermediate to high mass. More observations are needed
to confirm this.  If true, it could imply that enhanced gas
accretion and protostellar coalescence are important for high mass
stars in the densest environments (see more extensive reviews of
this point in Stahler, Palla, \& Ho 2000; Elmegreen 2004;
Shadmehri 2004).

\end{document}